\documentclass[aps, prd, reprint, amsfonts,
  amssymb, amsmath, showpacs, letterpaper]{revtex4-1}

\usepackage{braket, bm}

\usepackage{color,graphicx}

\usepackage{ulem}

\usepackage[caption=false]{subfig}

\begin{document}

\title{Radial acceleration relation from symmetron fifth forces}

\author{Clare Burrage}
\email{clare.burrage@nottingham.ac.uk}
\affiliation{School of Physics and Astronomy, University of Nottingham, Nottingham NG7 2RD, United Kingdom}

\author{Edmund J. Copeland}
\email{edmund.copeland@nottingham.ac.uk}
\affiliation{School of Physics and Astronomy, University of Nottingham, Nottingham NG7 2RD, United Kingdom}

\author{Peter Millington}
\email{p.millington@nottingham.ac.uk}
\affiliation{School of Physics and Astronomy, University of Nottingham, Nottingham NG7 2RD, United Kingdom}





\begin{abstract}
We show that the radial acceleration relation for rotationally-supported galaxies may be explained, in the absence of cold dark matter, by a non-minimally coupled scalar field, whose fifth forces are partially screened on galactic scales by the symmetron mechanism. In addition, we show that sufficient energy is stored in the symmetron field to explain the dynamic stability of galactic disks.
\end{abstract}

\maketitle

\section{Introduction}

The physical origins of dark matter and dark energy remain open and challenging theoretical questions. Scalar-tensor theories of gravity provide a potential explanation for dark energy~\cite{Copeland:2006wr,Clifton:2011jh}. However, their associated fifth forces have not been seen in local tests of gravity. The coupling of the scalar degree of freedom to matter must therefore be fine tuned or the fifth force must be screened in the local environment. Screening by local matter density (see, e.g.,~Ref.~\cite{Joyce:2014kja}) can arise through modifications to the mass of the scalar fluctuations, as in chameleon theories; the scalar kinetic term, as in the Vainshtein mechanism; or the matter coupling. An example of the latter is the symmetron~\cite{Hinterbichler:2010es,Hinterbichler:2011ca}, whose vacuum expectation value (vev) responds to the background matter density.

In this article, we argue that the response of a symmetron field to the baryonic density of rotationally-supported galaxies can significantly impact their dynamics, providing an explanation for galactic rotation curves that does not require particle cold dark matter (CDM). Although the present study suggests some tension with constraints from local tests of gravity, we illustrate how this model naturally leads to the correlation between the observed centripetal accelerations and those estimated from the baryonic component alone~\cite{Sanders1990,Janz:2016nwx,McGaugh:2016leg} --- the radial or mass-discrepancy acceleration relation. In addition, we show that the interactions between the baryons and the symmetron field can contribute sufficient potential energy to stabilize the galactic disk (see Ref.~\cite{Ostriker}).

Explanations for the observed correlation have also been suggested within MOdified Newtonian Dynamics (MOND)~\cite{Milgrom:2016uye} --- and by extension tensor-vector-scalar (TeVeS) theories (see, e.g., Ref.~\cite{Skordis:2009bf}) --- and generally covariant MOdified Gravity (MOG)~\cite{Moffat:2016ikl}. It has also been argued that this correlation is consistent with CDM if the dissipational collapse of the baryons is taken into account~\cite{Keller:2016gmw,Ludlow:2016qzh}. There is, of course, additional evidence for dark matter~\cite{Bertone:2016nfn}, e.g., large scale structure, the dynamics of galaxy clusters, measurements of weak lensing and observations of the Cosmic Microwave Background (CMB). Nevertheless, it is compelling that a model as simple as the one presented here can explain the observed rotation curves, whilst also providing an explanation for the stability of galactic disks.

The remainder of this article is organized as follows: After briefly introducing the symmetron model in Sec.~\ref{sec:model}, we proceed in Sec.~\ref{sec:rar} to describe how the spatial variation of the symmetron field profile can lead to flattened rotation curves that are consistent with the radial acceleration relation. In Sec.~\ref{sec:disk}, we illustrate how the symmetron interactions can stabilize the galactic disk. We present a numerical analysis of a sample of rotation curves in Sec.~\ref{sec:nums}, and our conclusions are given in Sec.~\ref{sec:conc}.

\section{Symmetron model}
\label{sec:model}

The symmetron model consists of a non-minimally coupled scalar field with Einstein-frame potential
\begin{equation}
\tilde{V}(\varphi)\ =\ \frac{1}{2}\bigg(\frac{\rho}{M^2}\:-\:\mu^2\bigg)\varphi^2\:+\:\frac{1}{4}\,\lambda\,\varphi^4\;,
\end{equation}
where $\mu^2>0$ and $\lambda>0$. The coupling to the non-relativistic energy density $\rho$ arises through the universal coupling of matter fields to the Jordan-frame metric $g_{\mu\nu}$, which is related to the Einstein frame metric $\tilde{g}_{\mu\nu}$ via the conformal transformation $g_{\mu\nu}=A^2(\varphi)\tilde{g}_{\mu\nu}$. The coupling function $A(\varphi)$ has the form
\begin{equation}
A(\varphi)\ =\ 1\:+\:\frac{\varphi^2}{2M^2}\:+\:\mathcal{O}\bigg(\frac{\varphi^4}{M^4}\bigg)\;.
\end{equation}
The scale $M$ determines the matter coupling strength.

As a result of the universal matter coupling, a unit test mass is subject to a fifth force (see, e.g., Ref.~\cite{Brax:2012nk})
\begin{equation}
\label{eq:symforce}
\vec{F}_{\rm sym}\: =\: -\,\vec{\nabla}\,\ln\,A(\varphi)\: \approx\: -\,\frac{\varphi}{M}\,\vec{\nabla}\,\frac{\varphi}{M}\quad (\varphi/M\ll 1)\;.
\end{equation}
In regions of low density, i.e.~$\rho/M^2  \ll\mu^2$, the model experiences spontaneous symmetry breaking, and the symmetron field acquires a nonzero vev $\varphi\approx \pm\,v=\pm\,\mu/\sqrt{\lambda}$. Any local spatial variation of the symmetron field then leads to an {\it unscreened} fifth force with coupling strength $v/M$. Instead, in regions of high density, i.e.~$\rho/M^2>\mu^2$, the minimum of the potential lies at the origin, the symmetry is restored, and $\varphi=0$. The coupling strength $\varphi/M$ therefore goes to zero, and the fifth force is {\it screened}.

\section{Radial acceleration relation}
\label{sec:rar}

We now describe how the spatial variation of the symmetron field, described in the preceding section and driven by the coupling to the baryonic density of the galaxy, leads to an additional acceleration consistent with the radial acceleration relation reported in~Ref.~\cite{McGaugh:2016leg}. This analysis of the SPARC data set~\cite{Lelli:2016zqa} showed that the observed centripetal accelerations ($g_{\rm obs}$) and those predicted from the baryonic component alone ($g_{\rm bar}$) follow the empirical relation
\begin{equation}
\label{eq:relrel}
g_{\rm obs}\ =\ \frac{g_{\rm bar}}{1-e^{-\sqrt{g_{\rm bar}/g_{\dag}}}}\ =\ g_{\rm bar}\:+\:\frac{g_{\rm bar}}{e^{\sqrt{g_{\rm bar}/g_{\dag}}}-1}\:\;,
\end{equation}
where $g_{\dag}=1.20\pm 0.02 \mathrm{(rand.)}\pm 0.24\mathrm{(sys.)}\!\times\! 10^{-10}\, {\rm ms^{-2}}$.

\begin{figure}
\centering
\includegraphics[scale=0.42]{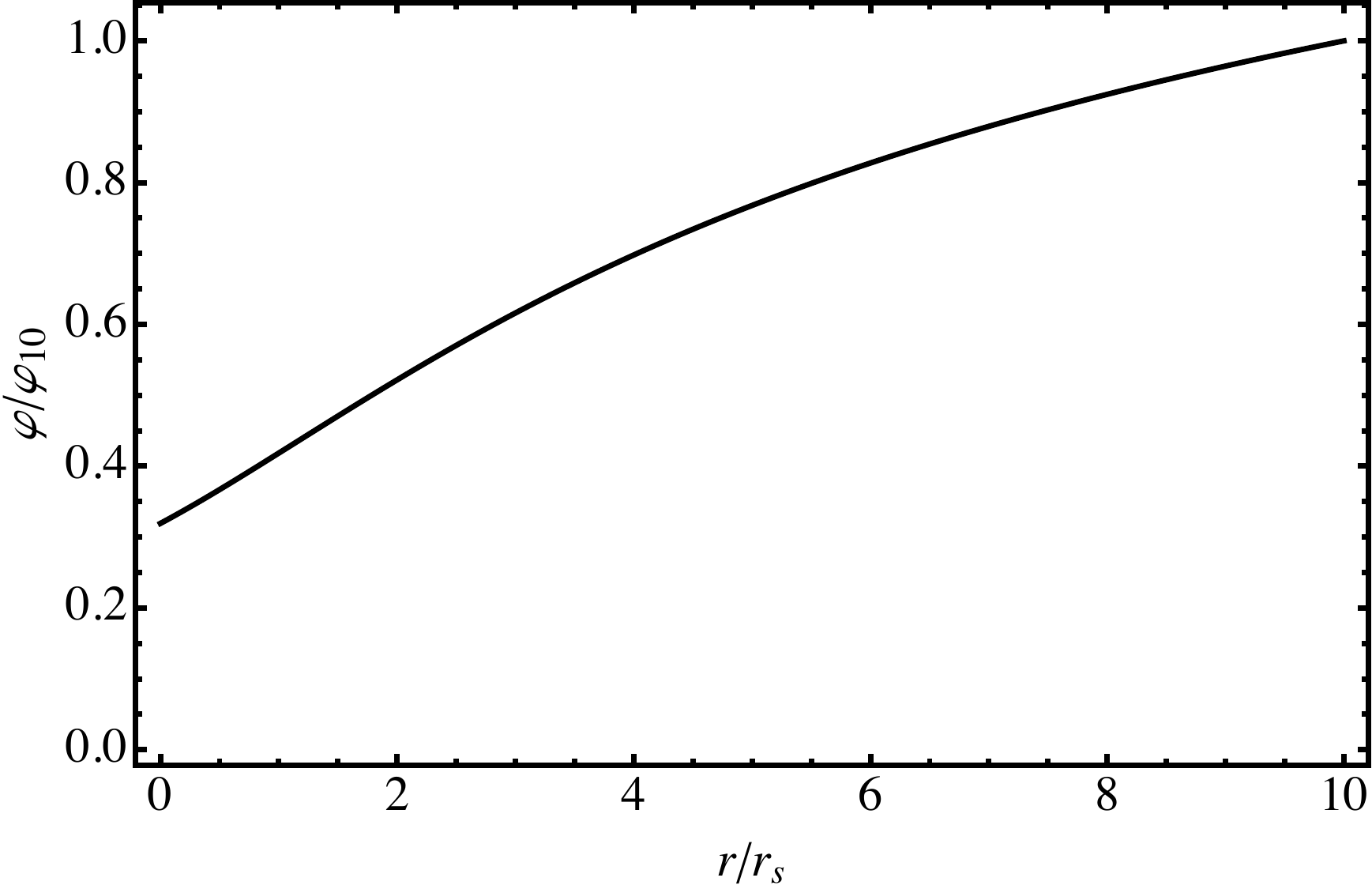}\vspace{-0.5em}
\caption{\label{fig1} The symmetron profile $\varphi$, required by Eq.~\eqref{eq:phiint} and normalized to the value of the field  at $r/r_s=10$ ($\varphi_{10}$) with boundary condition $\varphi_0/M=10^{-3}$.}
\end{figure}

Approximating the galaxies as thin disks (uniform in density over some height $h$), the symmetron force in Eq.~\eqref{eq:symforce} contributes a centripetal acceleration
\begin{equation}
\label{eq:match1}
g_{\rm sym}(r)\ =\ \frac{c^2}{2}\,\frac{\rm d}{{\rm d} r}\,\bigg(\frac{\varphi(r)}{M}\bigg)^{\! 2}\;,
\end{equation}
where $c$ is the speed of light. We neglect the restorative symmetron force normal to the plane of the disk, assuming the symmetron field to be approximately constant over the height of the disk. (By symmetry arguments, the field gradients normal to the disk must vanish as we approach the central plane of the disk.) The empirical correlation in Eq.~\eqref{eq:relrel} can therefore be explained if the profile of the symmetron field is such that
\begin{equation}
\label{eq:match2}
g_{\rm sym}(r)\ =\ \frac{g_{\rm bar}(r)}{e^{\sqrt{g_{\rm bar}(r)/g_{\dag}}}-1}\;,
\end{equation}
requiring
\begin{equation}
\label{eq:reqprof}
\bigg(\frac{\varphi}{M}\bigg)^{\!2}\ =\ \bigg(\frac{\varphi_0}{M}\bigg)^{\!2}\:+\:\frac{2}{c^2}\int_{0}^{r}{\rm d}r'\,\frac{g_{\rm bar}(r')}{e^{\sqrt{g_{\rm bar}(r')/g_{\dag}}}-1}\;,
\end{equation}
where $\varphi_0\equiv \varphi(0)$ is the value of the field at the origin.

Assuming an exponential disk profile for the surface mass density of the form
\begin{equation}
\label{eq:expdisk}
\Sigma(r)\ =\ \Sigma_0\,e^{-r/r_s}\;,
\end{equation}
the total mass within a radius $r$ is given by
\begin{align}
\mathcal{M}_{\rm bar}(r)\ &=\ \mathcal{M}_0\int_0^r\!\frac{{\rm d}r'}{r_s}\;\frac{r'}{r_s}\,e^{-r'/r_s}\nonumber\\& =\ \mathcal{M}_0\big[1-e^{-r/r_s}\big(1+\tfrac{r}{r_s}\big)\big]\;,
\end{align}
where $\mathcal{M}_0=2\pi r_s^2\Sigma_0$ is the total mass of the galaxy and $r_s$ is its scale length. Defining $x\equiv r/r_s$ and
\begin{equation}
f(x)\ \equiv\ \frac{f_0}{x} \big[1\:-\:e^{-x}\big(1+x\big)\big]^{\tfrac{1}{2}}\;,\quad f_0\ =\ \Big(\tfrac{G\mathcal{M}_0}{g_{\dag} r_s^2}\Big)^{\tfrac{1}{2}}\;,
\end{equation}
and using the fact that
\begin{equation}
g_{\rm bar}\ =\ \frac{G \mathcal{M}_{{\rm bar}}(r)}{r^2}\;,
\end{equation}
we see that the required field profile [Eq.~\eqref{eq:reqprof}] becomes
\begin{equation}
\label{eq:phiint}
\bigg(\frac{\varphi}{M}\bigg)^{\! 2}\ =\ \bigg(\frac{\varphi_0}{M}\bigg)^{\!2}\:+\:2\,\frac{g_{\dag}r_s}{c^2}\!\int_{0}^x\!{\rm d}x'\;\frac{f^2(x')}{e^{f(x')}-1}\;,
\end{equation}
where $f_0\approx 5$ for a galaxy with a mass and scale length comparable to the Milky Way ($\mathcal{M}_0\approx6\times10^{11}\ {\rm M}_{\odot}$ and $r_s\approx 5\ {\rm kpc}$). Figure~\ref{fig1} shows this profile as a function of $r/r_s$, normalized to its value at 10 scale lengths ($\varphi_{10}$). The integral in Eq.~\eqref{eq:phiint} is not bounded as $x \to \infty$, but this is not a problem, since the identification in Eq.~\eqref{eq:match2} need only hold out to a finite radius.

\begin{figure*}
\vspace{-1em}
\centering
\subfloat[][]{\includegraphics[scale=0.245]{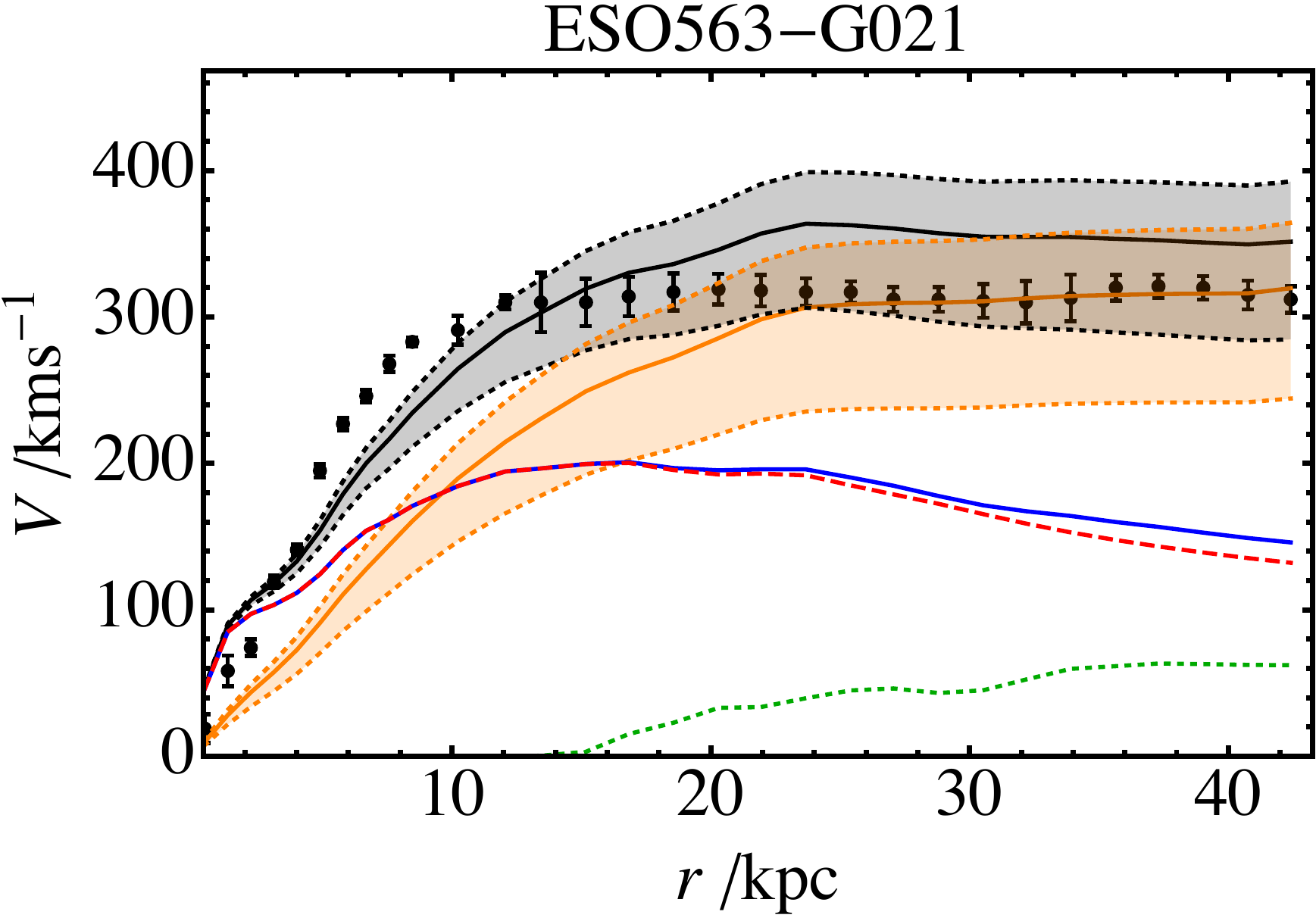}} \ \subfloat[][]{\includegraphics[scale=0.245]{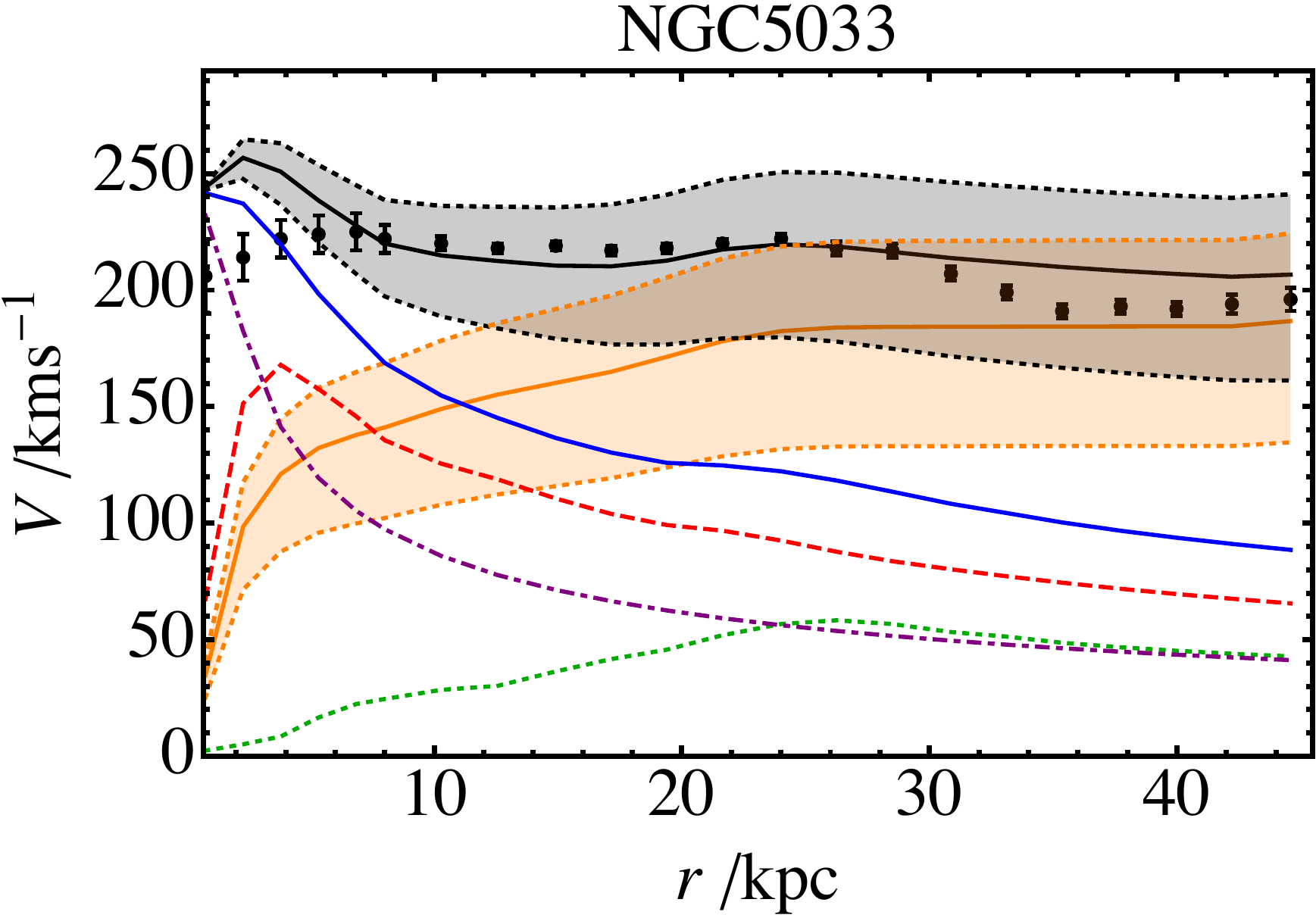}} \ \subfloat[][]{\includegraphics[scale=0.245]{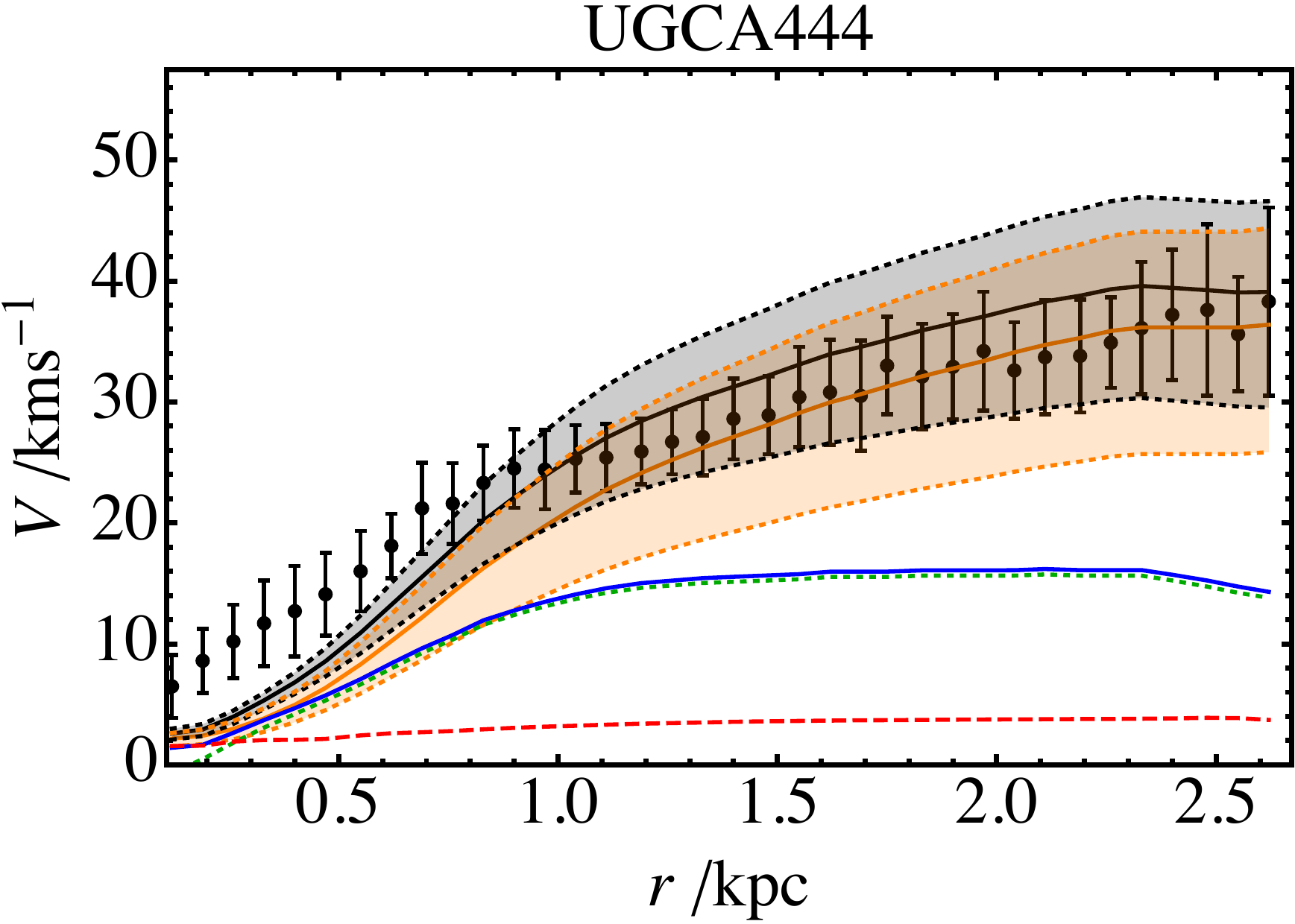}} \ \subfloat[][]{\includegraphics[scale=0.25]{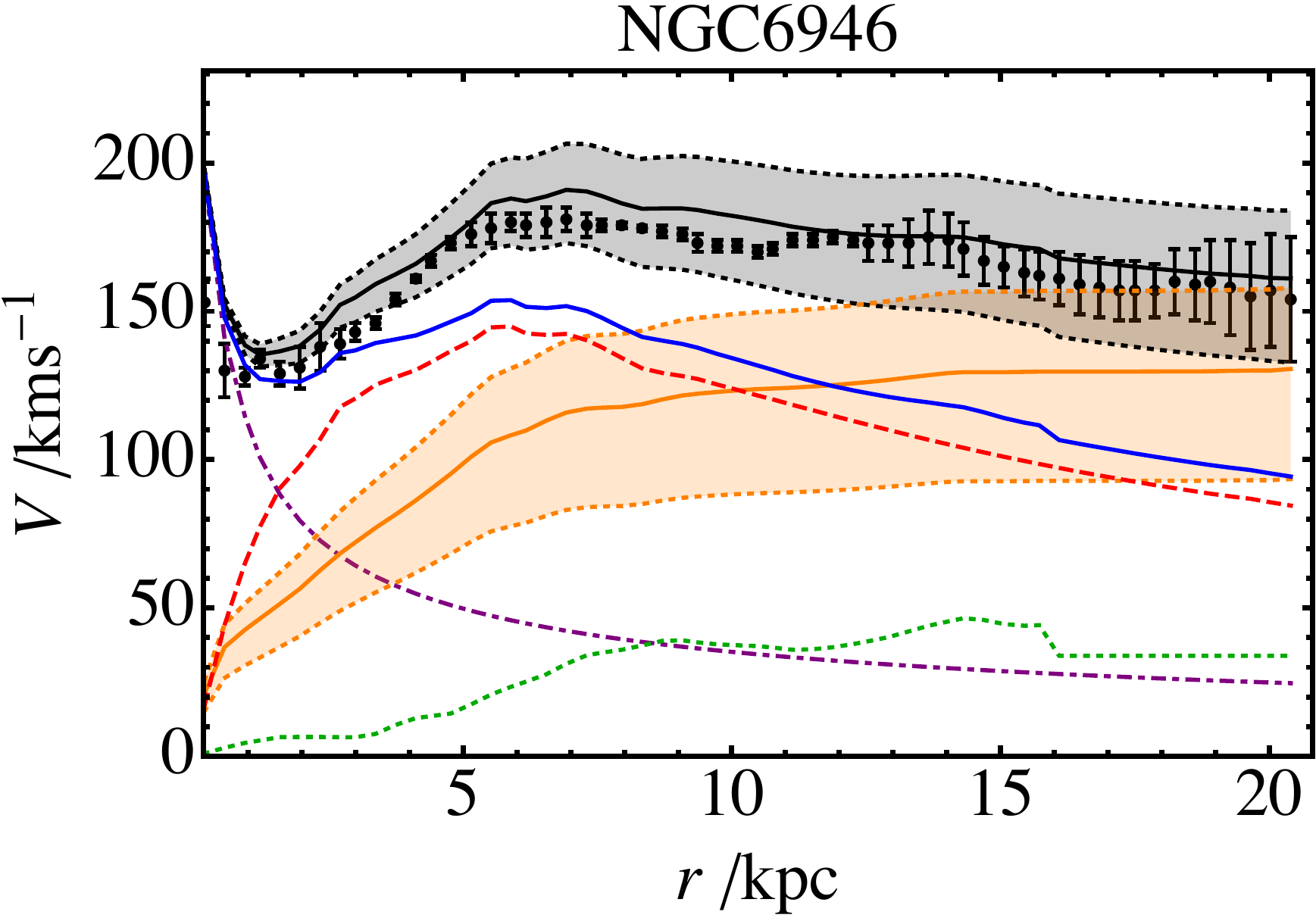}}\\\vspace{-0.75em}
\subfloat[][]{\includegraphics[scale=0.245]{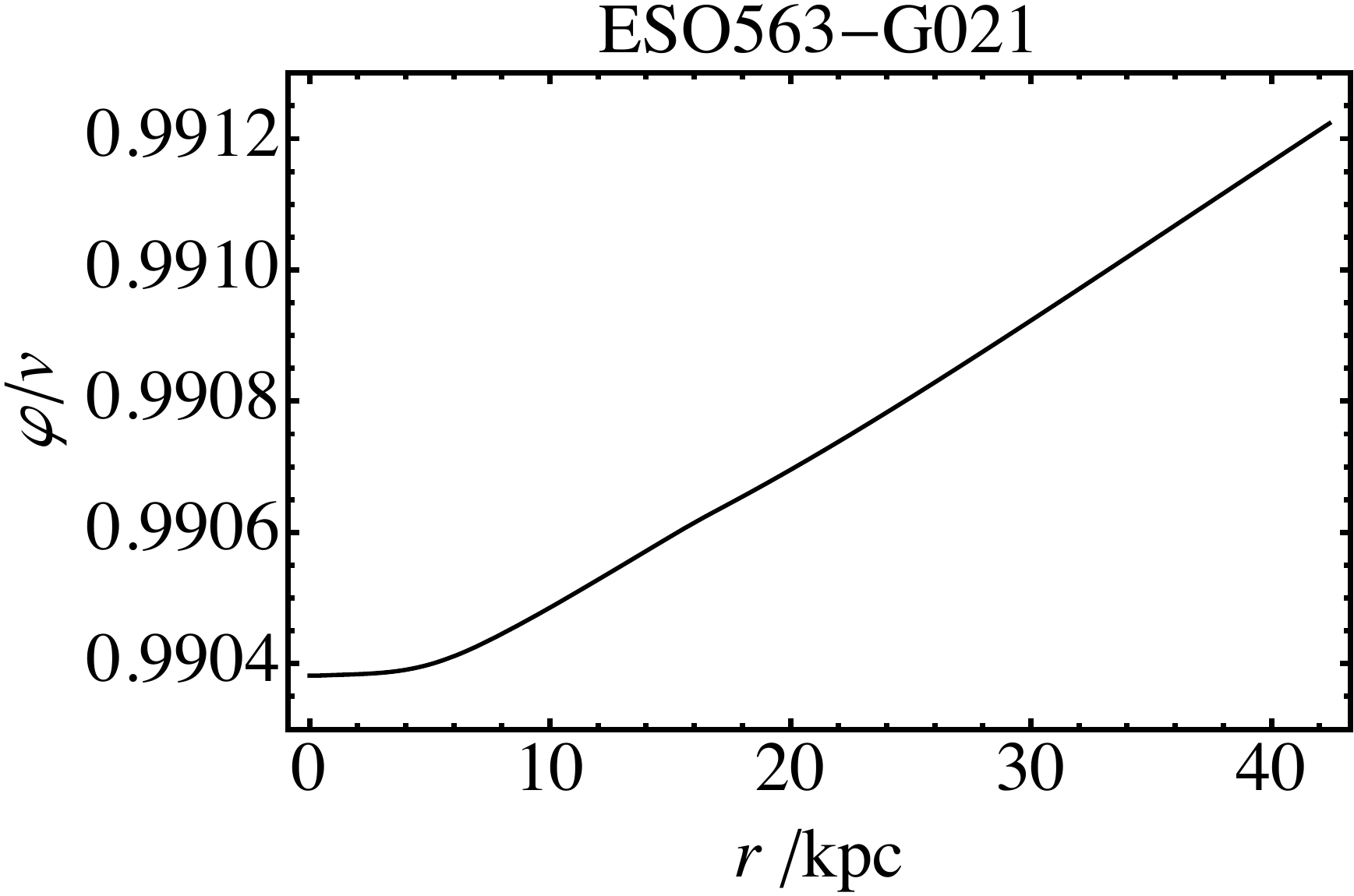}} \ \subfloat[][]{\includegraphics[scale=0.245]{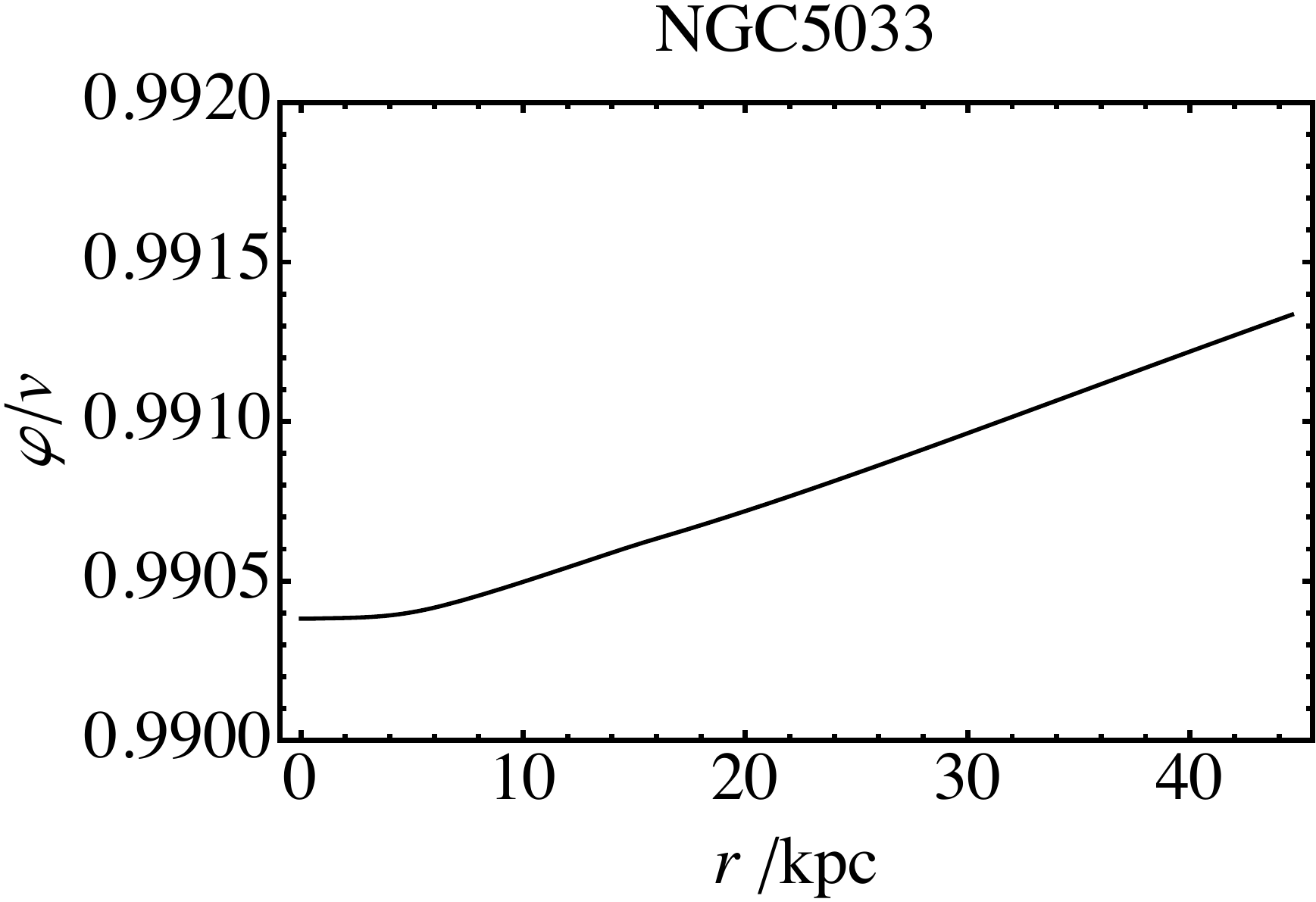}} \ \subfloat[][]{\includegraphics[scale=0.245]{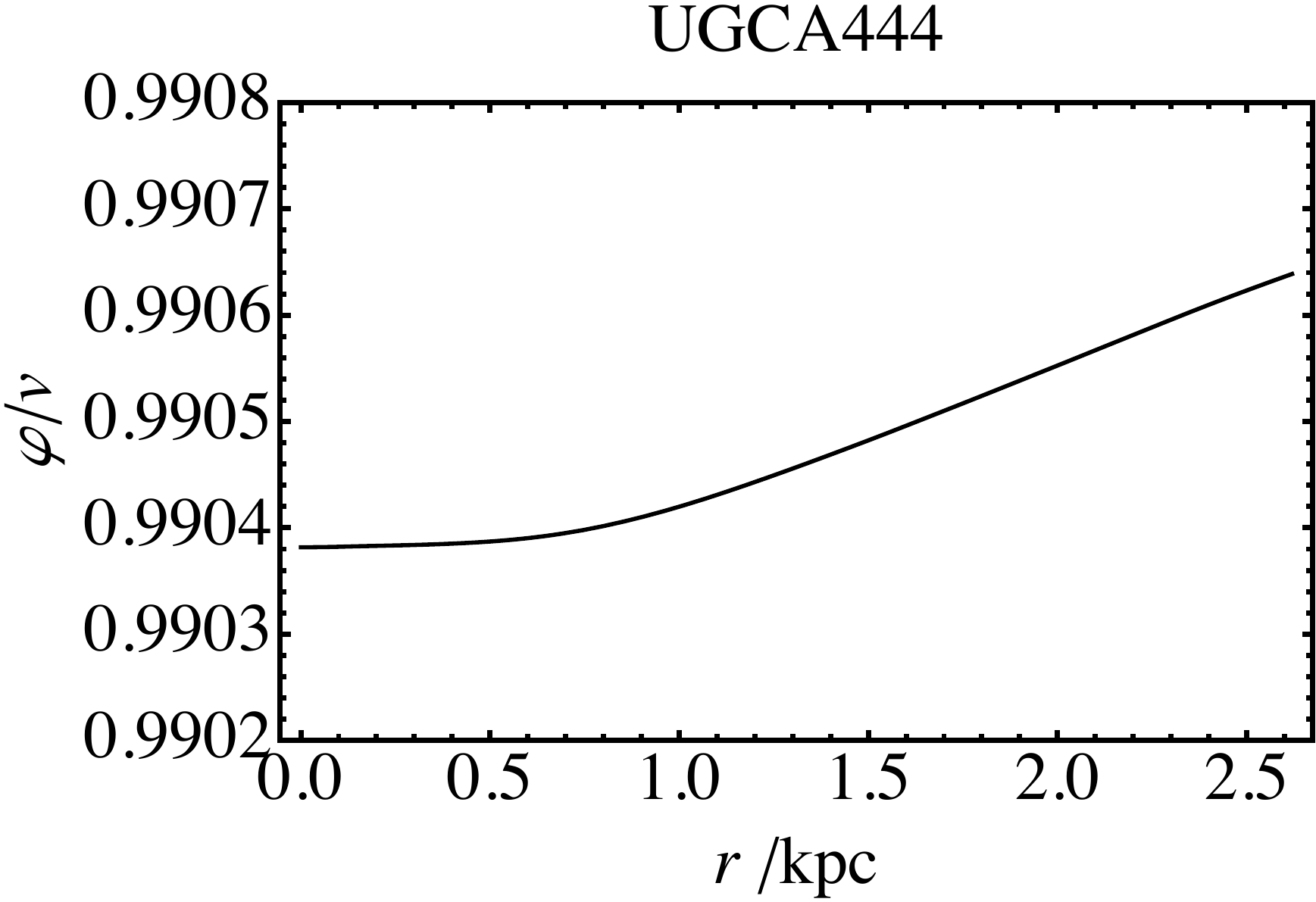}} \ \subfloat[][]{\includegraphics[scale=0.245]{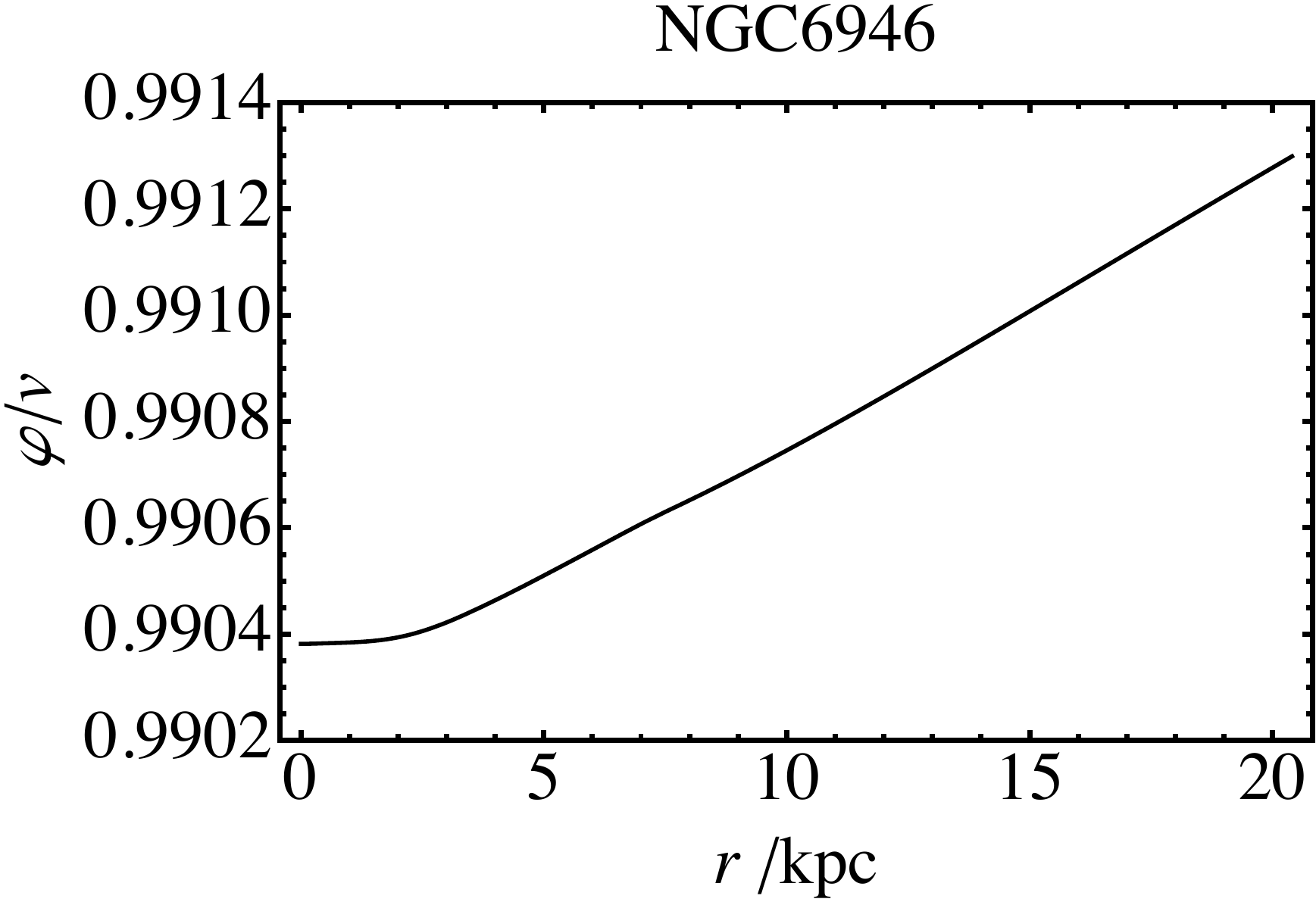}}\\\vspace{-0.75em}
\subfloat[][]{\includegraphics[scale=0.245]{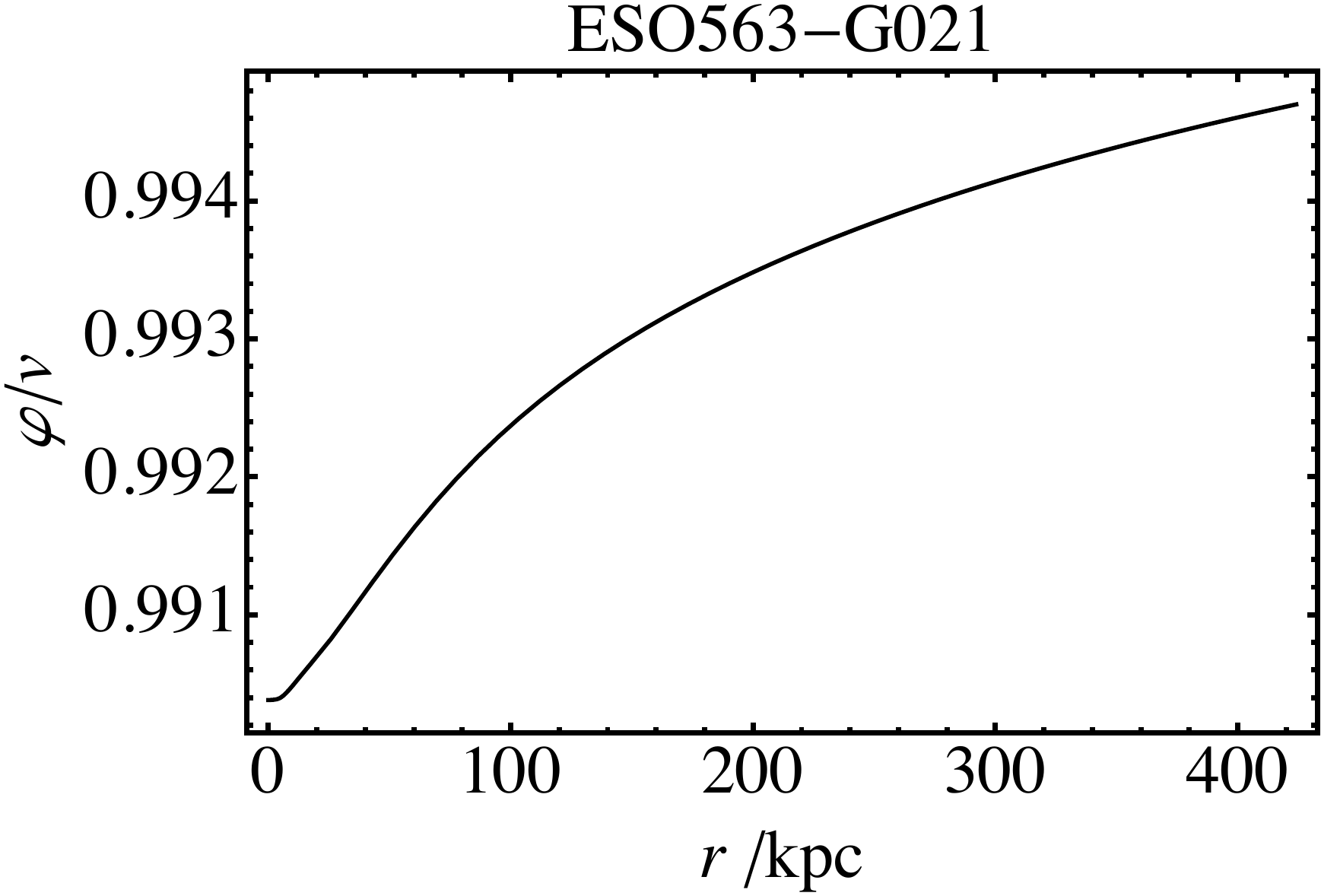}} \ \subfloat[][]{\includegraphics[scale=0.245]{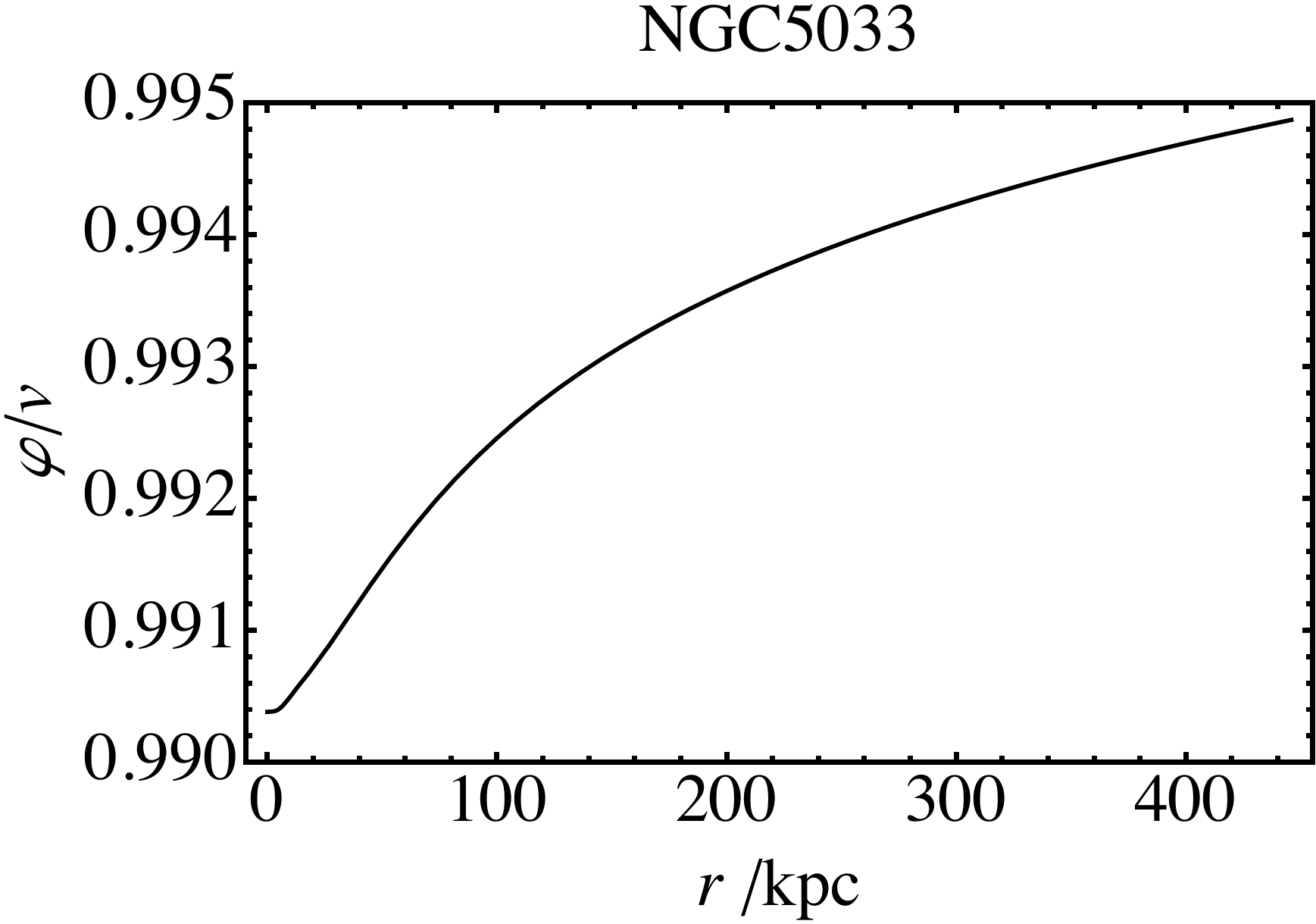}} \ \subfloat[][]{\includegraphics[scale=0.245]{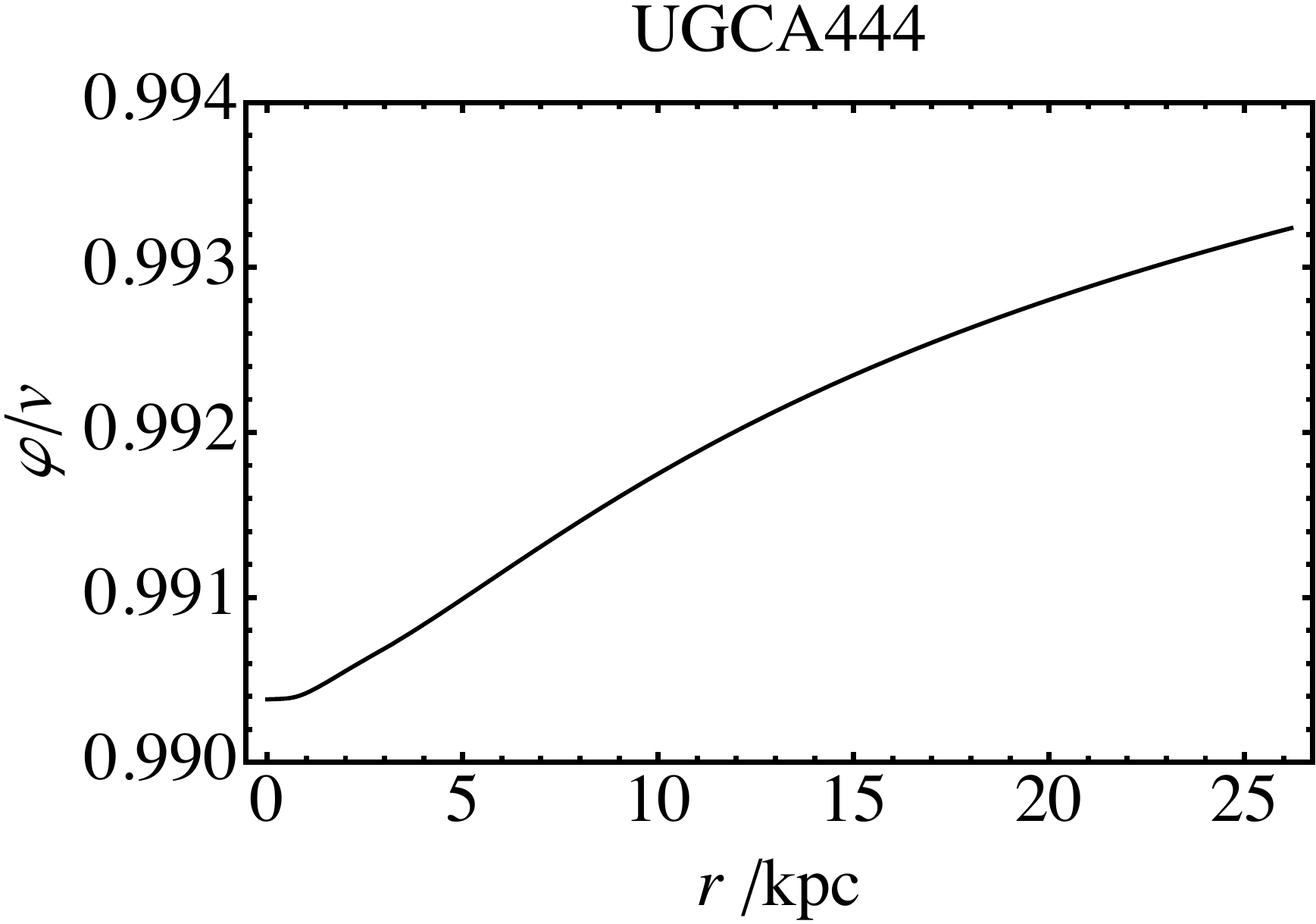}} \ \subfloat[][]{\includegraphics[scale=0.245]{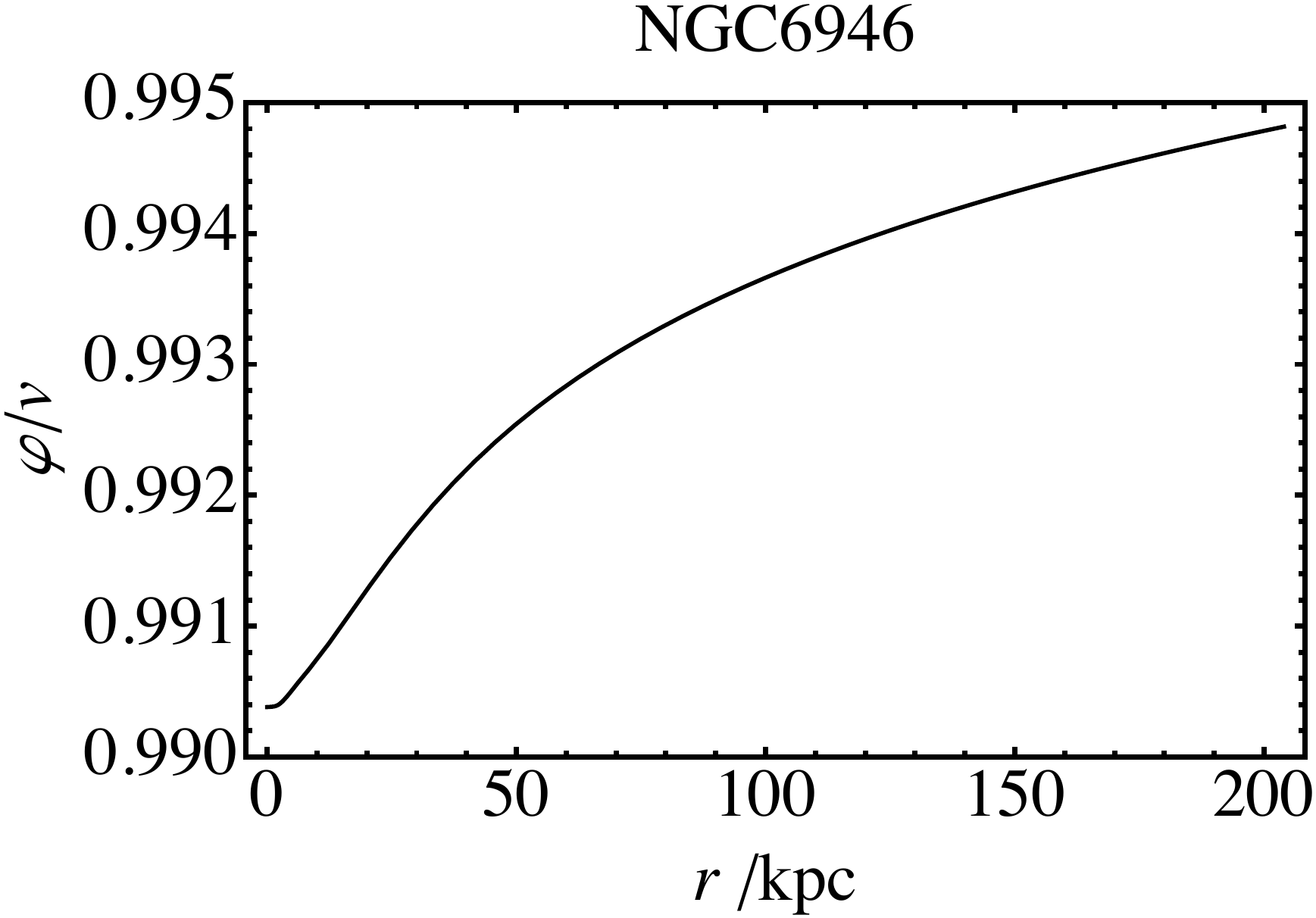}}\vspace{-0.75em}
\caption{\label{fig:profs} Example rotation curves for $M=M_{\rm Pl}/10$ and $\bar{\rho}_0 = 1\ {\rm M}_{\odot}\,{\rm pc}^{-3}$, $v/M= 1/150$, and $\mu=3\times10^{-39}\ {\rm GeV}$: (a) disk, (b) bulge and (c) gas dominated, and (d) comparable disk and bulge components. Black points: observed rotation velocities and corresponding error bars taken from the SPARC data set~\cite{Lelli:2016zqa}. Solid black: total prediction, including the symmetron component. Solid orange: symmetron contribution. Shaded bands indicate 50\% variation in $\bar{\rho}_0/M^2$. Solid blue: baryon-only prediction. Red dashed: disk component. Green dotted: gas component. Purple dot-dashed: bulge component. Figures (e)--(h) and (i)--(l) show the corresponding symmetron profiles over the observed data range and 10 times that range, respectively.}\vspace{-0.5em}
\end{figure*}

In the case of extended objects, the form of the symmetron force [Eq.~\eqref{eq:symforce}] is modified. For a star of radius $R_{\star}$, density $\rho_{\star}$ and mass $\mathcal{M}_{\star}$, the symmetron force per unit mass is
\begin{equation}
\vec{F}_{\rm sym}\ =\ -\,4\pi\,g_{\star}(\varphi)\,\vec{\nabla}\,\frac{\varphi}{\mathcal{M}_{\star}}\;,
\end{equation}
where the coupling strength $g_{\star}(\varphi)$ is
\begin{equation}
g_{\star}(\varphi)\ =\ \big(\varphi\:-\:\varphi_{\star}\big)\,\frac{R_{\star}\big[m_{\star}R_{\star}\:-\:\tanh(m_{\star}R_{\star})\big]}{m_{\star} R_{\star}\:+\:m_{{\rm gal}}R_{\star}\tanh(m_{\star} R_{\star})}\;.
\end{equation}
Here, $\varphi_{\star}$ is the value of the symmetron field at the centre of the star and $m_{\star({\rm gal)}}$ is the mass of the symmetron inside (outside) the star:
\begin{equation}
m_{\star({\rm gal})}^2\ =\ \begin{cases}
\frac{\rho_{\star({\rm gal})}}{M^2}\:-\:\mu^2\;, &\rho_{\star({\rm gal})}\ >\ \mu^2M^2\;,\\
2\big(\mu^2\:-\:\frac{\rho_{\star({\rm gal})}}{M^2}\big)\;, &\rho_{\star({\rm gal})}\ <\ \mu^2M^2\;.
\end{cases}
\end{equation}
The stars respond as point-like test masses, and Eq.~\eqref{eq:symforce} is exactly recovered, when $m_{\star} R_{\star}\ll 1$, $m_{\rm gal} R_{\star}\ll 1$ and $\varphi_{\star}\to 0$. This holds for the present case, where the symmetron Compton wavelengths internal and external to the star ($l\propto 1/m_{\star({\rm gal})}$) are larger than the stellar radii.

The symmetron force will also appear in the equations of hydrostatic equilibrium describing pressure-supported systems (cf.~Ref.~\cite{Milgrom:1984ij}), potentially explaining the observed velocity dispersions in, e.g., elliptical galaxies. The precise behaviour of the additional force depends upon the particular matter distribution and, in contrast to MOND, there is therefore no {\it a priori} reason for the effective acceleration scale ($g_{\dag}$) to be common to rotationally- and pressure-supported systems. This may explain the observed deviations of this acceleration scale (by a factor of a few). In addition, the effective lensing mass may be increased by including disformal couplings (see, e.g.,~Ref.~\cite{Khoury:2014tka}). We also remark that the ``kink-kink" interactions of the symmetron profiles, as well as the response of the symmetron field to the change in the gas distribution, may produce an offset between the stellar and ``dark matter'' components in colliding systems (see Ref.~\cite{Robertson:2016qef}), as has recently been observed in Abell 3827~\cite{Taylor:2017ipx}.

\section{Disk stability}
\label{sec:disk}

It is known that the baryonic component alone is insufficient to stabilize the disks of galaxies to barlike modes~\cite{Ostriker} and that this can be remedied by the presence of spherical dark matter halos. In what follows, we will show that the energy stored in the symmetron field can have a similar stabilizing effect.

Assuming an exponential disk profile and velocities given by the radial acceleration relation, the total rotational kinetic energy of the baryonic component $T$ and its potential energy due to Newtonian gravity $U$ are
\begin{subequations}
\begin{gather}
T\ \approx\ 4\,\mathcal{M}_0\,g_{\dag}r_s\quad (f_0\approx 5)\;,\\
U\ \approx\ -\:\frac{G\mathcal{M}_0^2}{2\,r_s}\;.
\end{gather}
\end{subequations}
The contribution to the Newtonian potential from the symmetron is negligible. However, additional potential energy results from the direct coupling between the baryons and the symmetron:
\begin{equation}
E_{\varphi}\ =\ \int\!{\rm d}V\;\frac{\rho\,\varphi^2}{2M^2}\ \approx\ \frac{\mathcal{M}_0}{2}\bigg(\frac{v}{M}\bigg)^2\!\int_0^{\infty}\!{\rm d}x\;x\,e^{-x}\,\bigg(\frac{\varphi}{v}\bigg)^2.
\end{equation}
The remaining integral, which we denote by the dimensionless parameter $\alpha$, is of order unity if the symmetron field does not vary from its vev significantly, i.e.~$\varphi\sim v$. The ratio of the total rotational kinetic energy of the baryons to the magnitude of their total potential energy 
is therefore given by
\begin{equation}
t\ \equiv\ \frac{T}{|U+E_{\varphi}|}\ \approx\ \bigg|-\:\frac{f_0^2}{8}\:+\:\frac{\alpha}{8\,g_{\dag}r_s}\bigg(\frac{v}{M}\bigg)^{\!2}\bigg|^{-1}\;.
\end{equation}

In order to ensure stability of the galactic disk, we require $t\lesssim 0.1376$~\cite{Ostriker}, thereby constraining
\begin{equation}
\label{eq:stability}
\frac{v}{M}\ \gtrsim\ \frac{4\times10^{-3}}{\sqrt{\alpha}}\;.
\end{equation}
In the next section, we will see that the symmetron fifth force can provide sufficient modification of the centripetal acceleration to flatten galactic rotation curves, whilst at the same time remaining consistent with this bound.

\begin{figure}
\centering
\subfloat[][observed versus baryonic]{\includegraphics[scale=0.448]{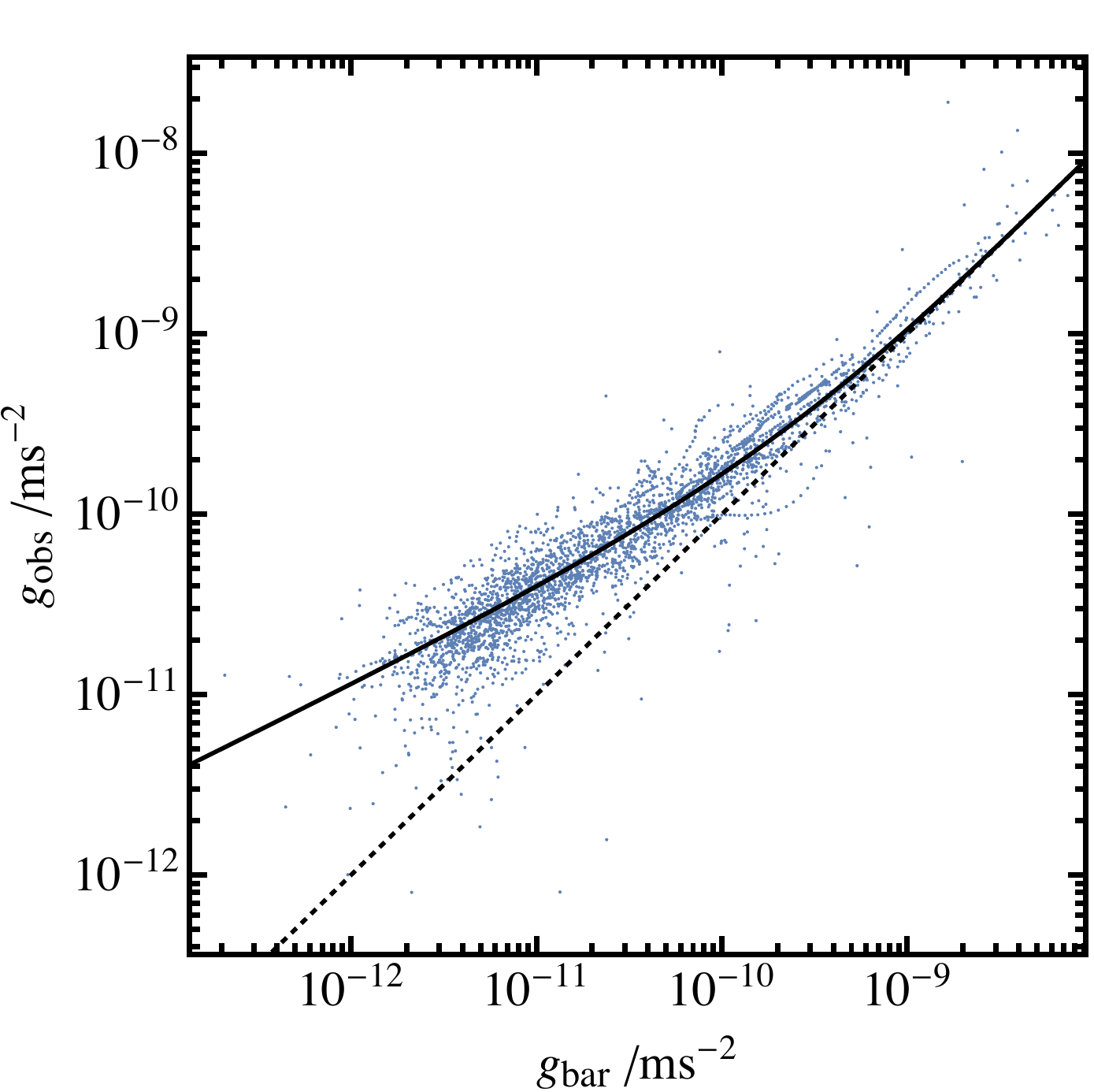}}\\
\subfloat[][symmetron prediction versus observed]{\includegraphics[scale=0.448]{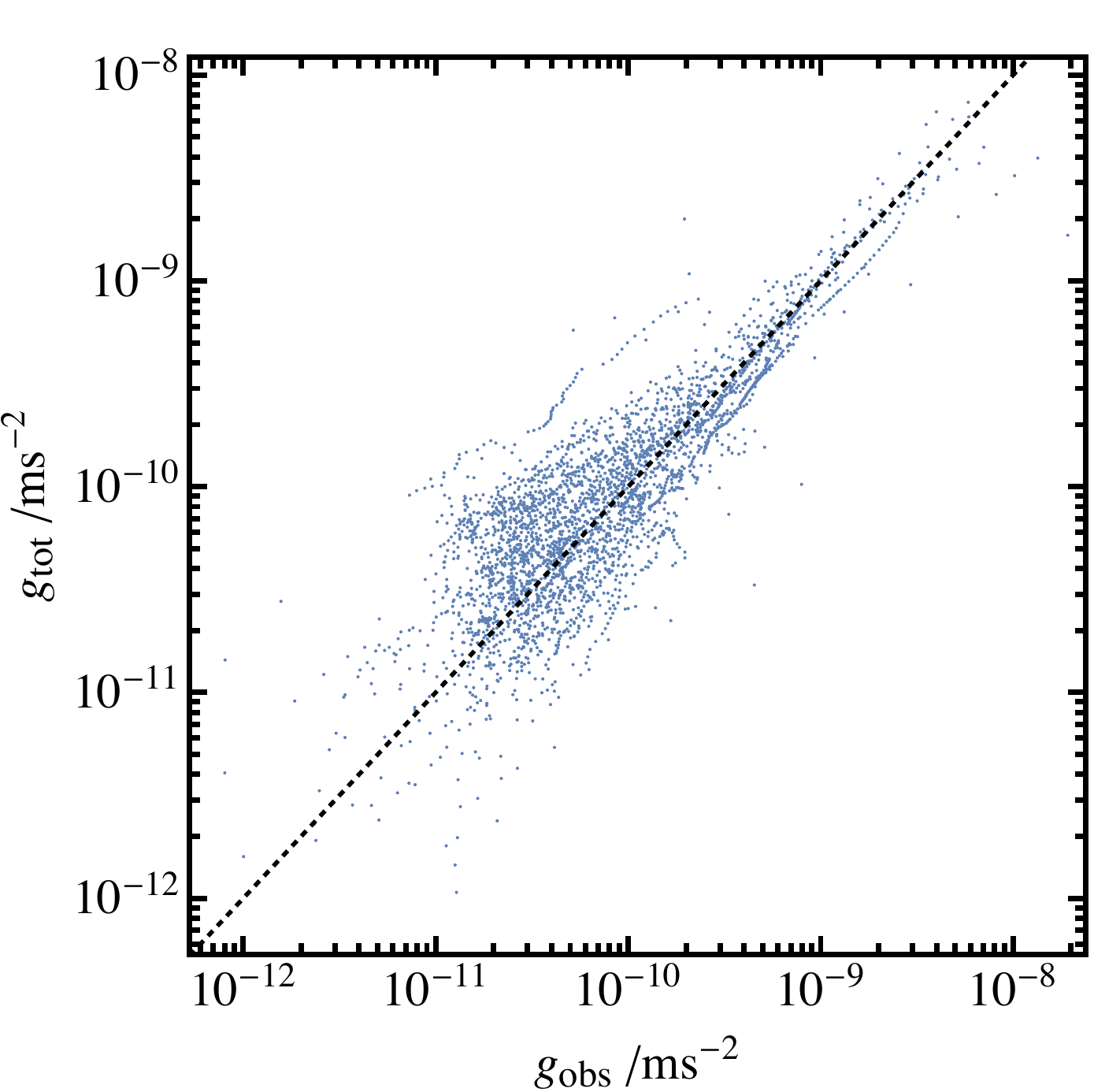}}\\
\subfloat[][symmetron prediction versus baryonic]{\includegraphics[scale=0.448]{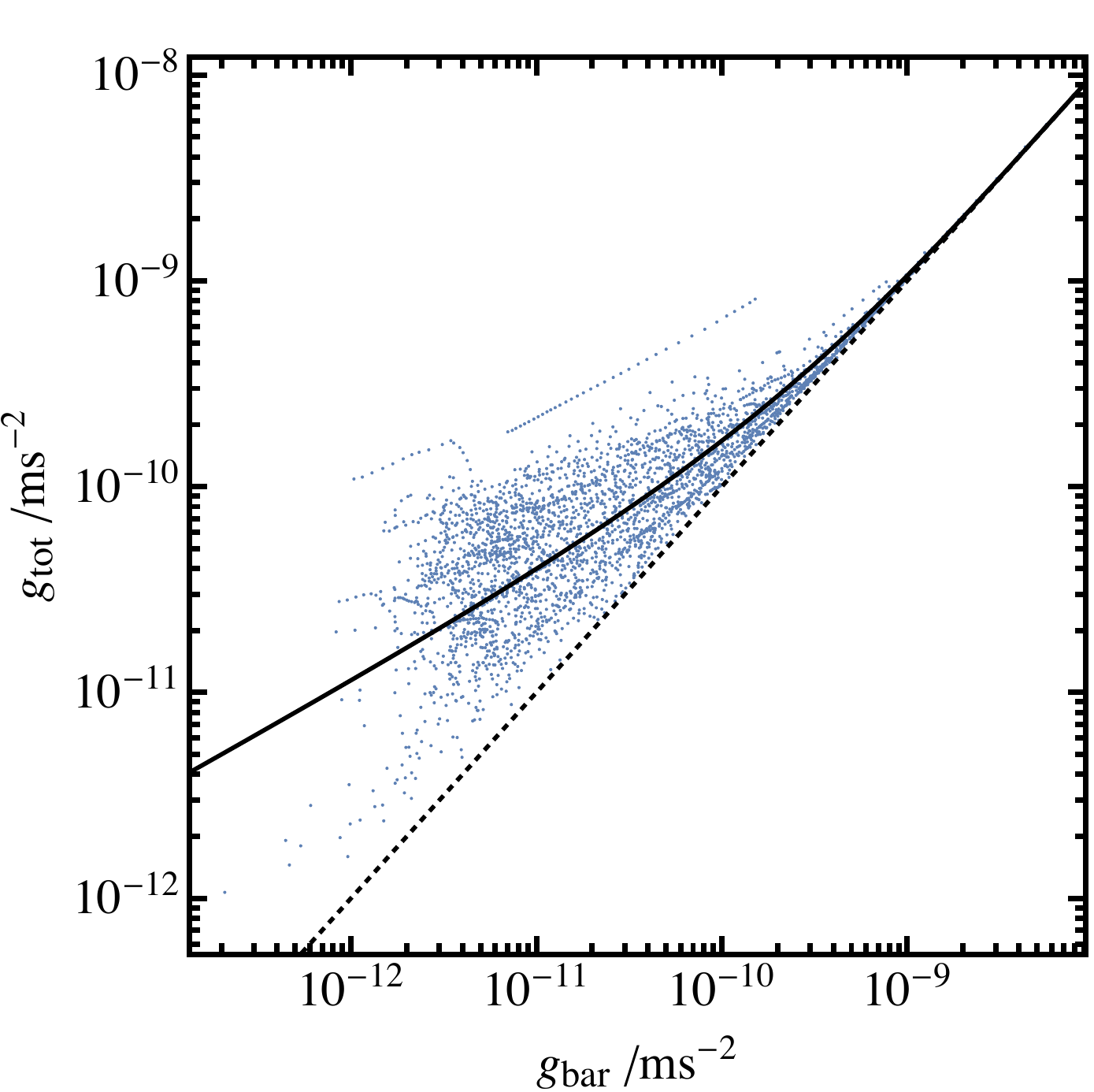}}
\caption{\label{fig:scatters}Acceleration parameters: (a) observed total ($g_{\rm obs}$) versus baryon-only prediction ($g_{\rm bar}$), cf.~Ref.~\cite{McGaugh:2016leg}; (b) predicted total acceleration for the symmetron model ($g_{\rm tot}$) versus observed total  ($g_{\rm obs}$); and (c) predicted acceleration  ($g_{\rm tot}$) versus baryon-only prediction ($g_{\rm bar}$). The solid black lines in (a) and (c) correspond to the radial acceleration relation [Eq.~\eqref{eq:relrel}].}
\end{figure}

\section{Numerical analysis}
\label{sec:nums}

We turn now to a numerical analysis of the symmetron field profiles and resulting rotation curves for a sample of galaxies in the SPARC data set.

We reconstruct the baryonic density profile from the SPARC mass models, assuming disk and bulge mass-luminosity relations of $\Upsilon_{\rm dis}=0.5\ {\rm M}_{\odot}/{\rm L}_{\odot}$ and $\Upsilon_{\rm bul}=0.7\ {\rm M}_{\odot}/{\rm L}_{\odot}$, as in Ref.~\cite{McGaugh:2016leg}:
\begin{equation}
\rho(r)\ \propto \ \frac{G\mathcal{M}_{\rm bar}'(r)}{r}\ =\ g_{\rm bar}(r)\:+\:2\,V_{\rm bar}(r)\,V'_{\rm bar}(r)\;.
\end{equation}
The radial derivatives (indicated by $'$) are estimated using a finite difference method. The density profile is extrapolated beyond the data range by fitting an exponential disk profile to the combined disk and gas components and a de Vaucouleur profile to any bulge component. In order to deal with the galaxy-by-galaxy uncertainties in the mass-luminosity relations and density profiles perpendicular to the disk, we make the coarse approximation that the average effective density to which the symmetron responds is constant over the SPARC sample, introducing the parametrization
\begin{equation}
\mu^2_{\rho}(r)\ \equiv \ \frac{\bar{\rho}_0}{\bar{\rho}}\,\frac{\rho(r)}{M^2}\;,
\end{equation}
where $\bar{\rho}$ is the average of the baryonic density and $\bar{\rho}_0\sim1\ {\rm M}_{\odot}\,{\rm pc}^{-3}$ sets the scale of the effective density.

Working in cylindrical coordinates and assuming an approximately separable solution (appropriate when $h\mu\ll 1$), the radial equation for the symmetron field around an isolated galaxy takes the form
\begin{equation}
\frac{1}{r}\,\frac{{\rm d}}{{\rm d}r}\bigg(r\,\frac{\rm d}{{\rm d}r}\,\varphi\bigg)\:-\:\mu_{\rho}^2(r)\,\varphi\:+\mu^2\,\varphi\:-\:\lambda\,\varphi^3\ =\ 0\;,
\end{equation}
subject to the boundary conditions $\varphi'(0)=0$ and $\varphi(r)|_{r\to\infty}=v$. Under this approximate separability, gradients perpendicular to the disk contribute an additional uncertainty on $\mu^2_{\rho}(r)$.

We solve for the symmetron profile over a finite range $[r_{\rm min},r_{\rm max}]$ using Mathematica's \texttt{NDSolve} routine. We take $r_{\rm min}\sim 0$ and $r_{\rm max}=120\,r_s$. Assuming an exponentially-decaying density profile, the asymptotic behaviors of the solution are
\begin{equation}
\varphi(r)\ \approx\ \begin{cases} A\,I_0\Big(\sqrt{\mu^2_{\rho}(0)-\mu^2}\,r\Big)\;,& \quad r\ \sim\ 0\;,\\
v\:-\:B\,K_0\Big(\sqrt{2\,\mu^2}\,r\Big)\;,& \quad r \ \gg\ r_s\;,
\end{cases}
\end{equation}
for $\mu_{\rho}(0)>\mu$ and $r_s\mu\ll1$, where $I_0$ and $K_0$ are the zeroth-order modified Bessel functions of the first and second kinds. The boundary conditions at $r_{\rm min}$ and $r_{\rm max}$ can therefore be specified independent of the unknown constants $A$ and $B$ as follows:
\begin{subequations}
\begin{gather}
\frac{\varphi'(r_{\rm min})}{\varphi(r_{\rm min})}\ =\ \frac{I_0'\big(\sqrt{\mu^2_{\rho}(0)-\mu^2}\,r_{\rm min}\big)}{I_0\big(\sqrt{\mu^2_{\rho}(0)-\mu^2}\,r_{\rm min}\big)}\;,\\
\frac{\varphi'(r_{\rm max})}{\varphi(r_{\rm max})-v}\ =\ \frac{K_0'\big(\sqrt{2\mu^2}\,r_{\rm max}\big)}{K_0\big(\sqrt{2\mu^2}\,r_{\rm max}\big)}\;.
\end{gather}
\end{subequations}

Figure~\ref{fig:profs} shows four examples of the rotation curves and symmetron profiles in good agreement with the data. These include one disk-dominated [Figs.~\ref{fig:profs} (a), (e) and (i)], one bulge-dominated [Figs.~\ref{fig:profs} (b), (f) and (j)], one gas-dominated [Figs.~\ref{fig:profs} (c), (g) and (k)], and one with comparable bulge and disk components [Figs.~\ref{fig:profs} (d), (h) and (l)]. The parameters of the model were taken to be $M=M_{\rm Pl}/10$ (for $\bar{\rho}_0=1\ {\rm M}_{\odot}\,{\rm pc}^{-3}$ and where $M_{\rm Pl}$ is the reduced Planck mass), $v/M= 1/150$ and $\mu=3\times10^{-39}\ {\rm GeV}$.  Shaded bands correspond to 50\% variation in $\bar{\rho}_0/M^2$. The parameters were chosen so as to remain in the weakly non-linear regime, $r_s^2\mu^2\ll 1$, and are consistent with disk stability [see Eq.~\eqref{eq:stability}] for reasonable values of $\sqrt{\alpha}\gtrsim 3/5$. The mass $\mu > \sqrt{3}H_0 M_{\rm Pl}/M$ (cf.~Ref.~\cite{Hinterbichler:2010es}), where $H_0$ is the present-day Hubble constant, ensures that the symmetry is broken in the cosmological vacuum today.

In the weakly non-linear regime, the galaxies are unscreened at all radii, placing the present analysis in tension with Solar System constraints (see Refs.~\cite{Hinterbichler:2011ca} and~\cite{Burrage:2016xzz}). Observations of nearby distance indicators, i.e.~cepheids, water masers and tip of the red giant branch (TRGB) stars, also indicate that these objects must be largely screened within dwarf galaxies~\cite{Jain:2012tn}. We suggest that this tension may be lessened by moving to the strongly non-linear regime at smaller values of $M$ and larger values of $\mu$. In this case, the fifth force will be more strongly screened at our radius from the Galactic Centre, becoming fully unscreened only at larger radii (where more significant modifications to the dynamics are required). In addition, local variations of the symmetron profile within the galaxy will be enhanced. However, in this regime, the disparity between the galactic scale length $r_s$ and the symmetron Compton wavelength leads to a highly stiff and numerically challenging differential system. Even so, by keeping a comparable ratio of $\mu^2_{\rho}(0)/\mu^2$, one might continue to explain the rotation curves and disk stability. This tension may also be lessened by invoking additional screening, e.g., via the Vainshtein mechanism~(cf.~Ref.~\cite{Khoury:2014tka}).

The top two panels of Fig.~\ref{fig:scatters} show the observed velocities versus the baryon-only [Fig.~\ref{fig:scatters} (a)] (cf.~Ref.~\cite{McGaugh:2016leg}) and symmetron predictions [Fig.~\ref{fig:scatters} (b)] for the 153 galaxies~\cite{Endnote} analysed in Ref.~\cite{McGaugh:2016leg}. The symmetron force is always attractive and so no acceleration parameters are predicted below those inferred from the baryonic component [see Fig.~\ref{fig:scatters} (c)]. In addition, the baryon-only and symmetron predictions converge at high accelerations, since the screening of the fifth force is maximal towards the galactic centre. The scatter in the symmetron predictions at low accelerations is in part due to the uncertainty on the three-dimensional density. However, having not binned the data, the contributions of individual galaxies are visible. Each shows a similar correlation with the baryonic predictions up to some systematic scaling, which may have a physical origin. We emphasise that the present analysis treats each galaxy in isolation. In reality, the symmetron will be sensitive to the galaxy's local environment, providing an additional source of scatter.  Moreover, variation of $g_{\dag}$, e.g., with red-shift~\cite{Keller:2016gmw}, might be expected.

\section{Conclusions}
\label{sec:conc}

We have shown that the symmetron mechanism can explain galactic rotation curves and the stability of galactic disks. This alone does not eliminate the need for dark matter, and some tension with local tests of gravity remains, but it motivates further study of the intriguing alternative to the $\Lambda$CDM paradigm provided by symmetron-like mechanisms (see also~Refs.~\cite{Burrage:2016xzz, Damour:1994zq, Pietroni:2005pv, Olive:2007aj, Brax:2010gi, Brax:2011ja}). At the very least, we have illustrated how a non-minimally coupled scalar field could alter the estimated density of dark matter halos and, at best, we have provided a compelling explanation of rotation curves and disk stability that relies solely on density-driven spontaneous symmetry breaking.

\begin{acknowledgments}
This work was supported by STFC Grant No. ST/L000393/1 and a Royal Society University Research Fellowship. The authors would like to thank Alfonso Arag\'{o}n-Salamanca, Anne Green, Justin Khoury, Federico Lelli and Richard Massey for comments, suggestions and helpful discussions.
\end{acknowledgments}

\end{document}